\documentclass{JAIS}
\pdfoutput=1
\journal{JAIS-ID}
\vol{2021}

\received{14 December 2021}
\published{xx March 2022}

\def\be{\begin{equation}}
\def\ee{\end{equation}}
\def\bea{\begin{eqnarray}}
\def\eea{\end{eqnarray}}

\newcommand{\bi}{\bibitem}

\usepackage{caption}
\usepackage{multicol}
\usepackage{subcaption}
\usepackage{comment}
\usepackage{url}
\begin{document}

\title{A new versatile method for the reconstruction of scintillator-based muon telescope events}

\author{Raphaël Bajou\auno{1}, Marina Rosas-Carbajal\auno{1}, and Jacques Marteau\auno{2}}
\address{$^1$Université de Paris, Institut de Physique du Globe (CNRS UMR 7154), Paris, France}
\address{$^2$Université de Lyon UCBL, Institut de Physique des deux infinis (CNRS UMR 5822), Lyon, France}

\begin{abstract}
    This paper presents a new method to process the data recorded with muon telescopes. We have developed this processing method for the plastic scintillator-based hodoscopes located around the volcano La Soufrière de Guadeloupe, in the French Lesser Antilles, in order to perform muon radiographies of the lava dome region, strongly impacted by the volcanic hydrothermal activity.
    Our method relies on particle trajectory reconstruction, performing a fit of the recorded hits in the impacted scintillator bars using a Random Sample Consensus algorithm. This algorithm is specifically built to discriminate outlier points, usually due to noise hits, in the data. Thus, it is expected to significantly improve the signal/noise separation in muon track hits and to obtain higher quality estimates of the particles' incident trajectories in our detectors. The first analysis of the RANSAC-reconstructed events offers promising results in terms of average density maps.
    To illustrate the performances of this algorithm, we provide angular resolution and reconstruction efficiency estimates using a GEANT4 simulation of a telescope equipped with four detection matrices. In addition, we also show preliminary results from open-sky data recorded with such telescope at La Soufrière de Guadeloupe volcano.

\end{abstract}

\maketitle


\section{Introduction}
The intense hydrothermal activity occurring at La Soufrière de Guadeloupe, in the French Lesser Antilles, is a major concern for volcanologists at the Volcanological and Seismological Observatory of Guadeloupe (OVSG) and for the local population that lives nearby. Partly hosted within the andesitic lava dome, this hydrothermal system, continuously fueled with meteoric waters, is indeed responsible for both violent phreatic eruptions (last major event occurred in 1976-1977), and fast rock alteration due to hot and acid fluid circulation that worsens the risk of a partial volcano flank collapse. 

Since 2015, the deployment of particle trackers developed at IP2I, Lyon, to perform dynamic muon imaging of the lava dome structure, in addition to other geophysical methods such as seismic monitoring, has  allowed to increase the knowledge of the hydrothermal system dynamics (see \cite{legonidec2019}). 
Nowadays two new generation particle trackers equipped with 4 scintillator panels are installed at two different locations: South, shown on figure \ref{photo_snj}, and East of the volcano. 
This 4-panel configuration allows to increase precision on the 2D density radiography of the central part of the scanned region and offers two more 3-panels sub-configurations, which lead finally to three different muon images.
Previously acquired data by our group were based on 3-panel detectors, and processed using a direct straightness check on the scintillator bars hits in each panel. Here, we propose a more complex particle reconstruction algorithm, which takes into account the presence of outlier events and allows to improve the precision on the reconstructed trajectories.

\section{Detection technology}
\label{secDetector}
\begin{figure}[!ht]
    \begin{subfigure}[b]{.33\textwidth}
      \centering
        \includegraphics[width=\textwidth]{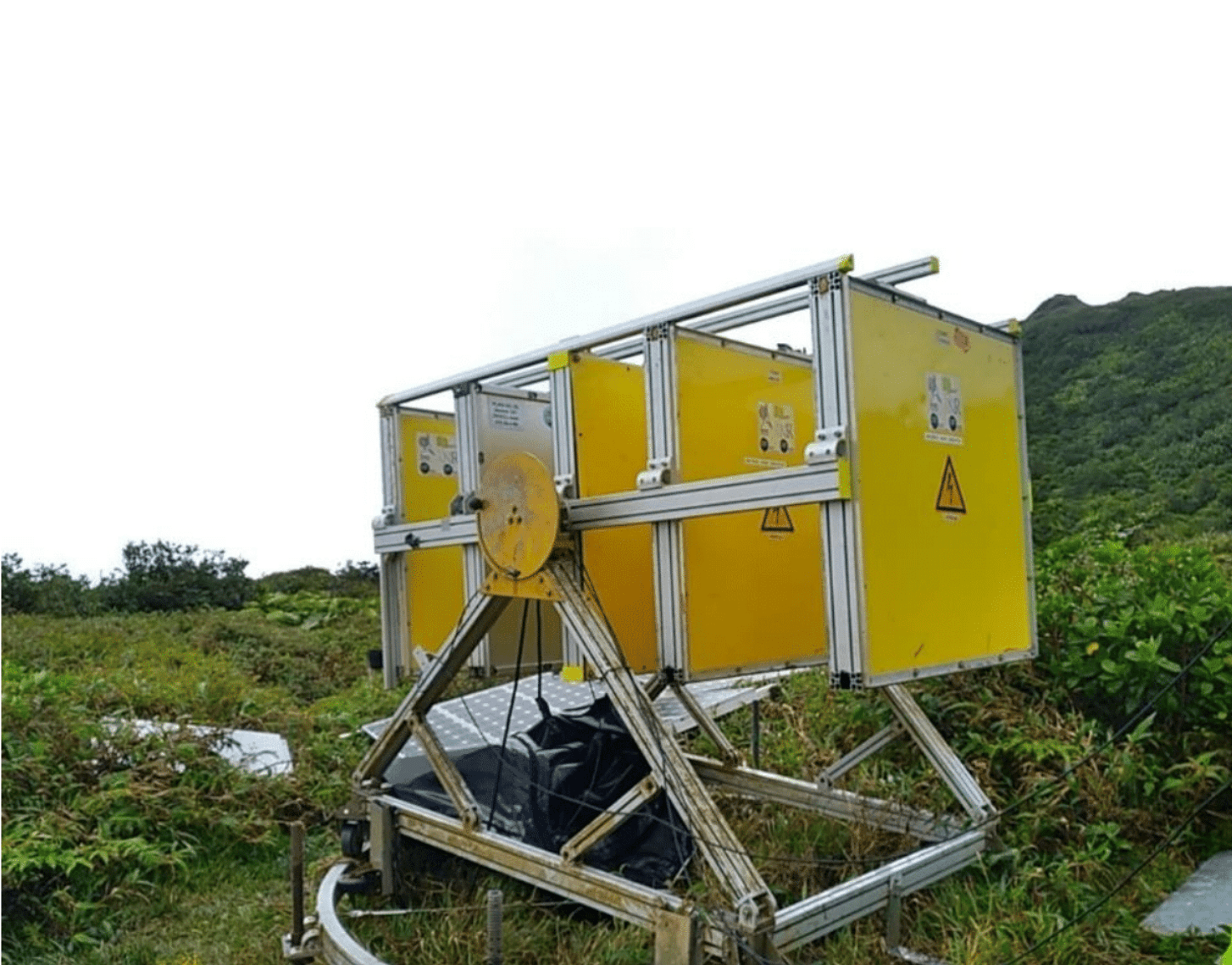}
        \caption{} 
        \label{photo_snj}
    \end{subfigure}
    \hfill
    \begin{subfigure}[b]{.33\textwidth}
      \centering
      \includegraphics[width=\textwidth]{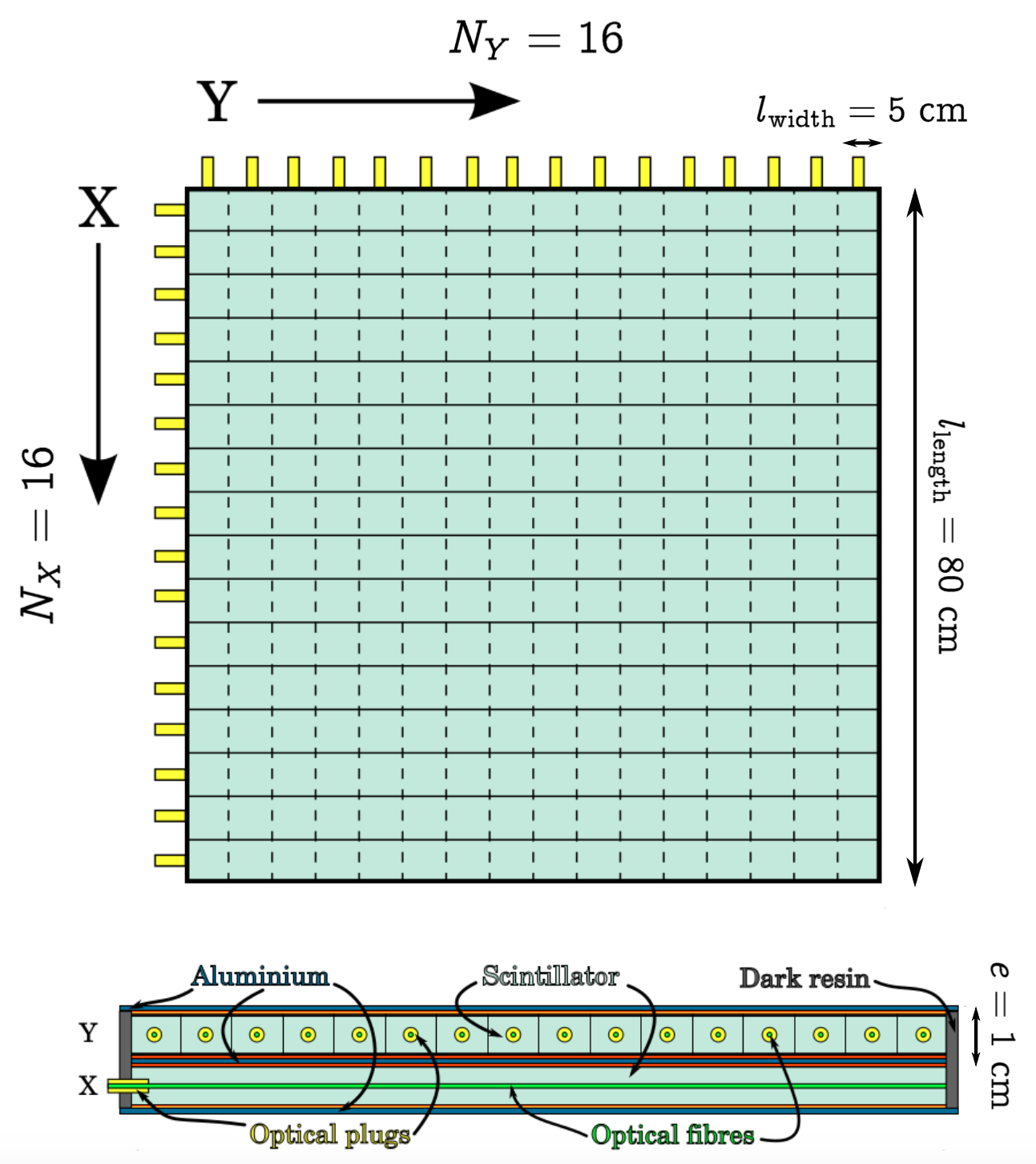}
      \caption{}
      \label{scheme_matrix}
    \end{subfigure}%
    \hfill
    \begin{subfigure}[b]{.33\textwidth}
      \centering
        \includegraphics[width=\textwidth]{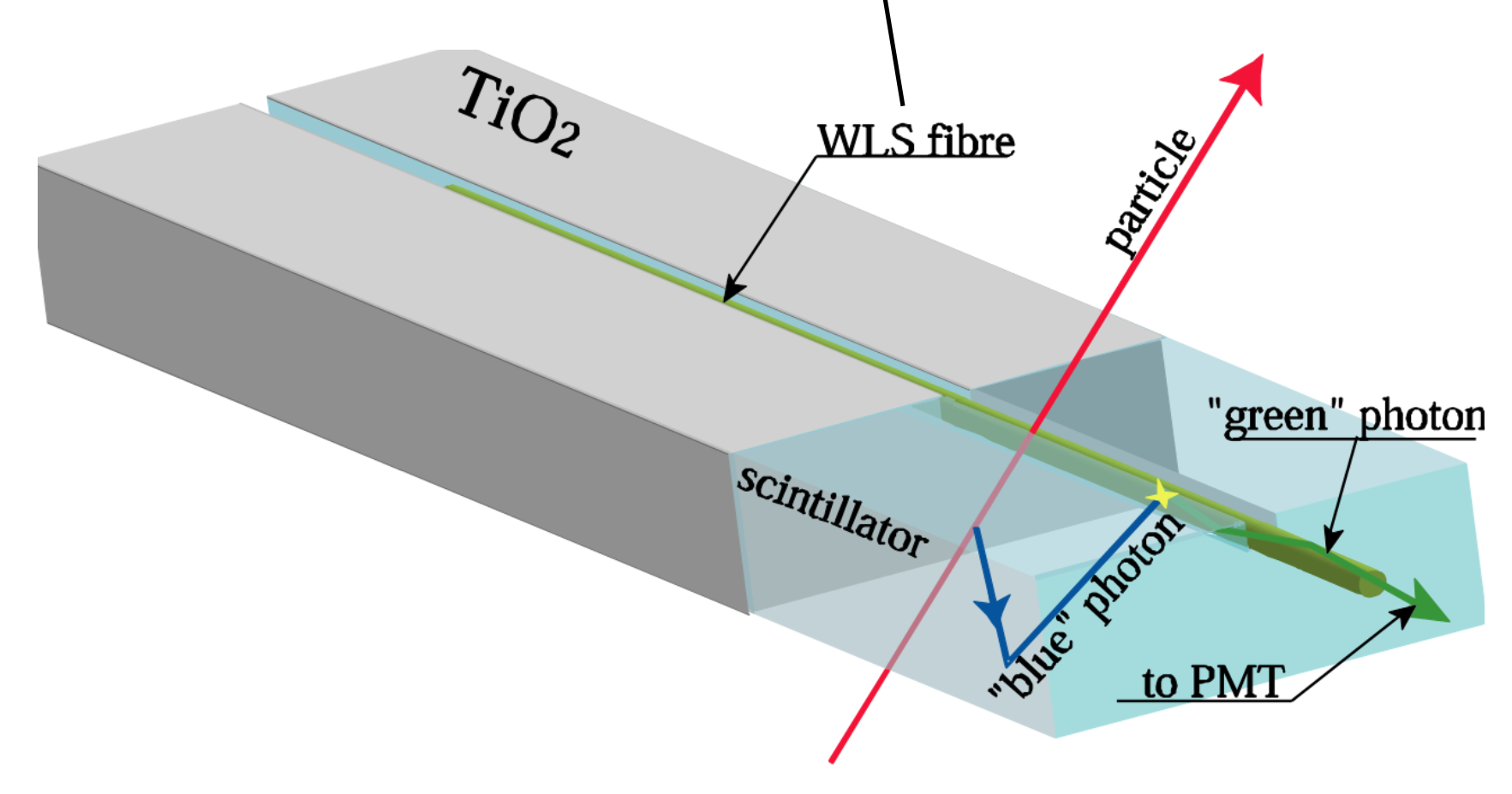}
        \caption{} 
        \label{optical_fiber}
    \end{subfigure}
\caption{(a) photo of the 4-panel detector \textit{Super Nain Jaune}, located near La Savane à mulets on the South flank of La Soufrière volcano; (b) a scheme of a scintillator matrix, adapted from \cite{lesparre2012}; (c) a scheme of a plastic scintillator bar, with its embedded WLS fiber, adapted from \cite{opera} }.
\label{detector}
\end{figure}

The detection panels are composed of an array of 16x16 plastic-scintillator bars of 5 cm width (detection area 80x80 $cm^{2}$)  (fig. \ref{scheme_matrix}). Each bar has its own Wavelength Shifting Fibers (WLS) embedded within the scintillator medium (fig. \ref{optical_fiber}), to collect the light emitted along the charged particle path, and guide the signal towards multi-anodes photomultipliers. 
In addition to the four detection matrices, a 100 mm thick lead shielding panel has been installed between the third and the fourth (rear) matrices, satisfying mechanical constraints on the telescope structure. This passive shielding allows to stop low-energy particles and also trigger electromagnetic showers inside our detection volume initiated by non-muonic events characterized with high hit multiplicity on the rear panel caused by secondary shower particles.
For more information about the detector assembly, we invite the reader to refer to \cite{lesparre2012} and \cite{jourde2013}; for details about the light and electronics readout system to \cite{marteau2010} and \cite{marteau2013}.

\newpage
\section{{Data processing and performances}}
\label{secProcess}
 Within telescope data, the detected muon events outgoing from the volcano are polluted with noise resulting from the electromagnetic component of atmospheric showers, and secondary electrons from interactions inside the detector, as well as an important contribution from low-energy muons scattered on volcano flanks \cite{gomez2017}. 
 To mitigate the noise contribution on radiographies, raw data are processed event-by-event with a preliminary filter on the minimum number of impacted detection panels, and hit multiplicity. Events with a multiplicity larger than 10 on one panel are rejected, to remove electromagnetic shower-like events triggered after shielding panel crossing. After filtering, to minimize the impact of the noisy hits on the measurement of the incident muon direction, we have chosen to reconstruct each track with a random sampling consensus (RANSAC) fit procedure. 
 \subsection{The RANSAC procedure}
 The RANSAC algorithm \cite{ransac} is based on an hypothesize and test iterative process.
 The process starts by randomly sampling within the input dataset, i.e. the hits XYZ coordinates, a subset of points of given size and fit an hypothesized line model to this subset. It then evaluates which hits of the whole dataset are consistent with this hypothesized model, computing a fit error, and counting the number of hits whose orthogonal distance to the model is below a certain distance threshold (set to one scintillator width).  Those hits forms the consensus set. After N iterations, the algorithm outputs the model that minimizes the fit error, and maximizes the size of the consensus set. The hits belonging to the consensus are tagged as 'inliers' and the rest of the dataset as 'outliers'. N is set so that the probability of getting a pure-signal consensus, i.e. only composed of true inliers, at the end of the process is 0.99.

 \subsection{{Numerical simulation}}
 \label{simu}
\begin{figure}[!ht]
    \centering
    \includegraphics[width=0.25\textwidth]{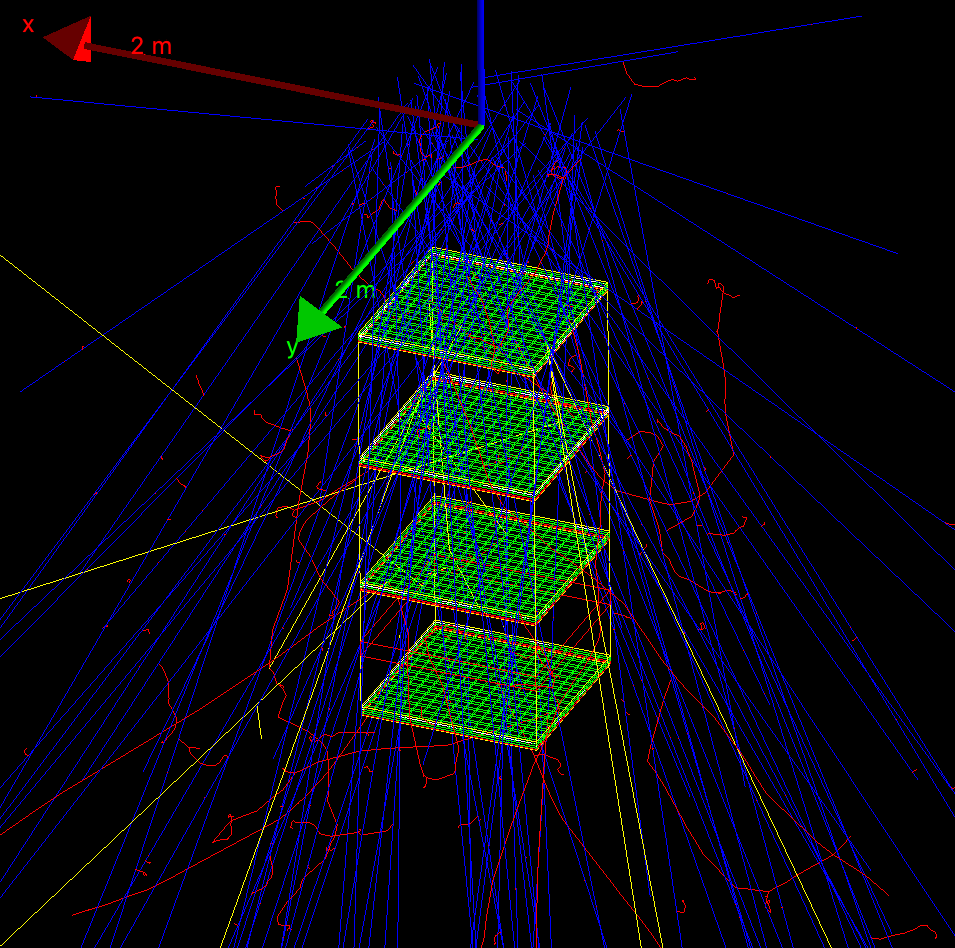}
    \caption{An event display of cosmic muons emitted from a plane surface above the front panel of the 4-panel detector using the CRY MC generator. The scintillator matrices are colored in green, the shielding parts are in red, and a 100-mm-thick lead panel is located between the third and the fourth panel.}
    \label{evd_mu_cry}
\end{figure}

To measure the performances of this RANSAC-based reconstruction, we have developed a dedicated GEANT4 \cite{geant4} simulation with our 4-panel detector geometry, illustrated on figure \ref{evd_mu_cry}, interfaced with the MC generator \textit{CRY} \cite{cry} to simulate the incident flux from atmospheric showers (muonic, hadronic and electromagnetic components). CRY appeared well-suited for processing performance tests, since it is very fast: $O(1 min)$ for $10^5$ MC events. 

\subsection{{Performances}}
\label{perf}
Using the simulation framework described in \ref{simu}, we first compared the RANSAC processing output of pure-muon and pure-electron samples of $10^6$ particles each.
We then measured, for the muon sample, the reconstruction efficiency and the angular resolution of the estimated trajectories. We compared these measurements with the results obtained from a simpler processing algorithm relying on a track straightness criteria, without any  trajectory fit involved. This former processing was based on the check of the alignment of the most energetic hit formed in each impacted panel, which is a significant bias since muons are minimum ionizing particles and therefore would not be expected to systematically form the highest energy deposit in each detection matrix. 

\subsubsection{Muon vs electron signal}
\begin{figure}[ht!]
    \begin{subfigure}[ht]{.45\textwidth}
      \centering
      \includegraphics[width=0.9\textwidth]{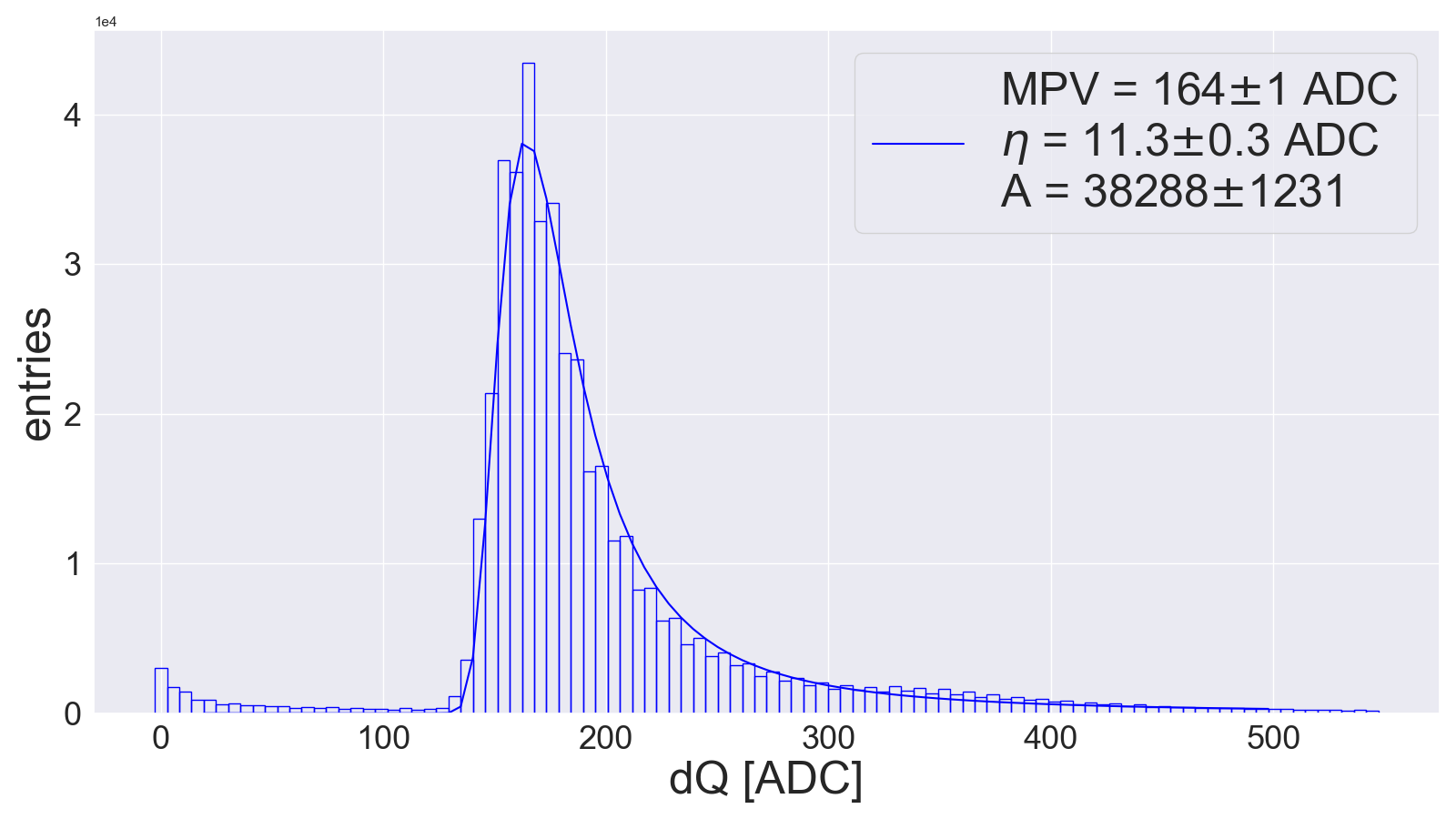}
      \caption{Charge signal for muons $dQ_{\mu}$ before processing}
      \label{dQ_mu_raw}
    \end{subfigure}%
    \hfill
    \begin{subfigure}[ht]{.45\textwidth}
      \centering
        \includegraphics[width=0.9\textwidth]{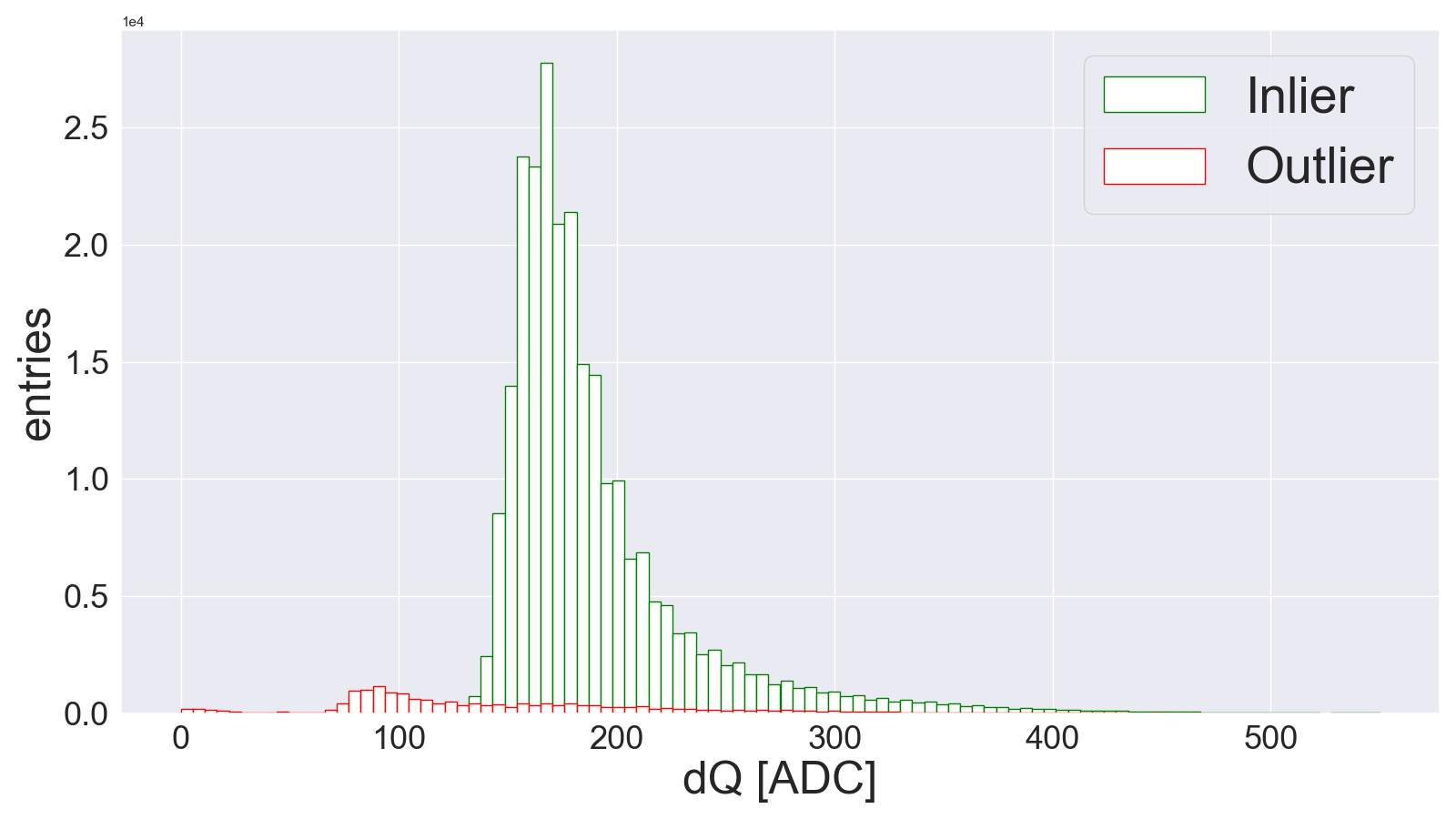}
        \caption{$dQ_{\mu}$ after processing} 
        \label{dQ_mu_ransac}
    \end{subfigure}
    
    \begin{subfigure}[ht]{.45\textwidth}
      \centering
      \includegraphics[width=0.9\textwidth]{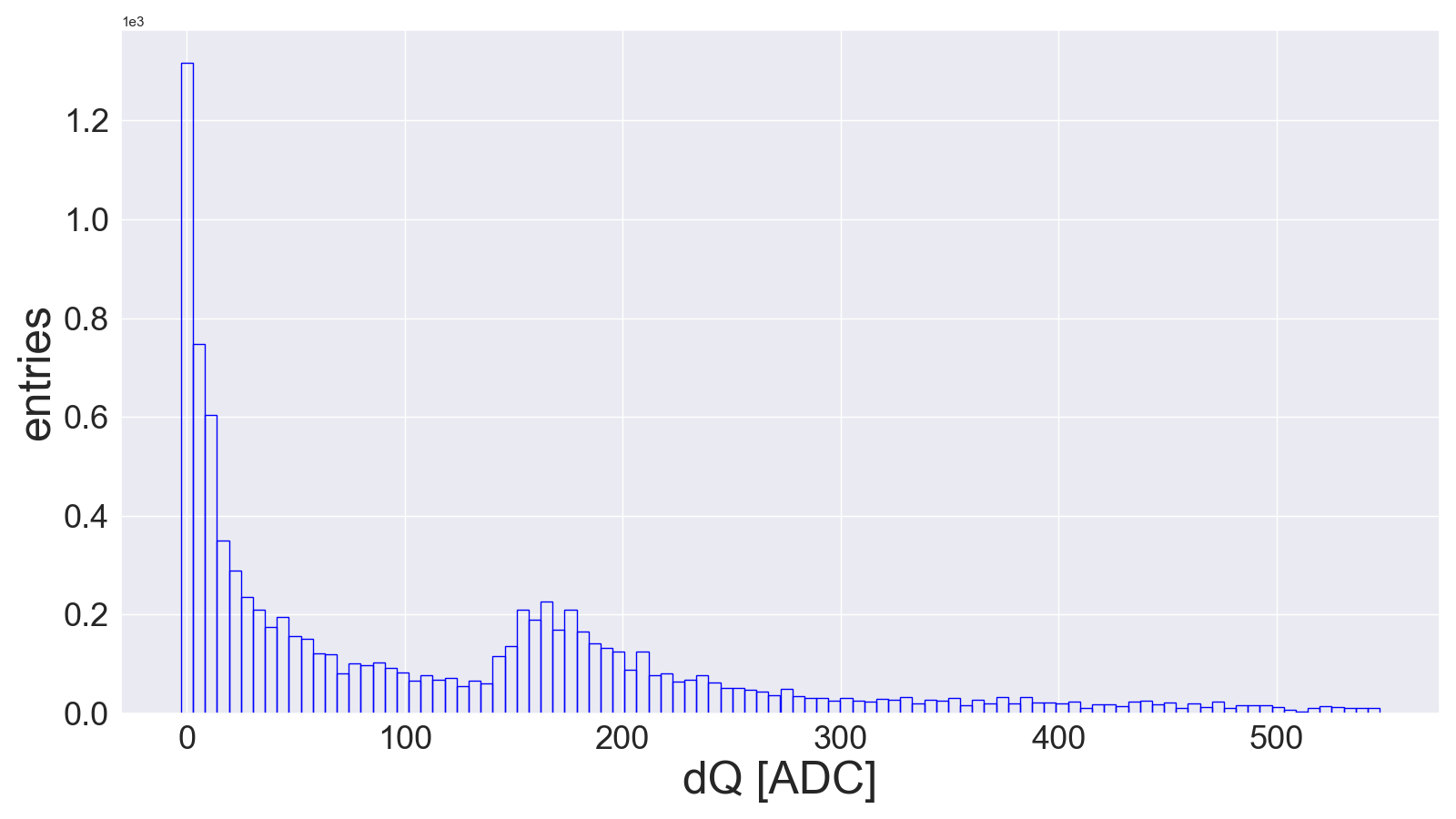}
      \caption{Charge signal for electrons $dQ_{e}$ before processing}
      \label{dQ_mu}
    \end{subfigure}%
    \hfill
    \begin{subfigure}[ht]{.45\textwidth}
      \centering
        \includegraphics[width=0.9\textwidth]{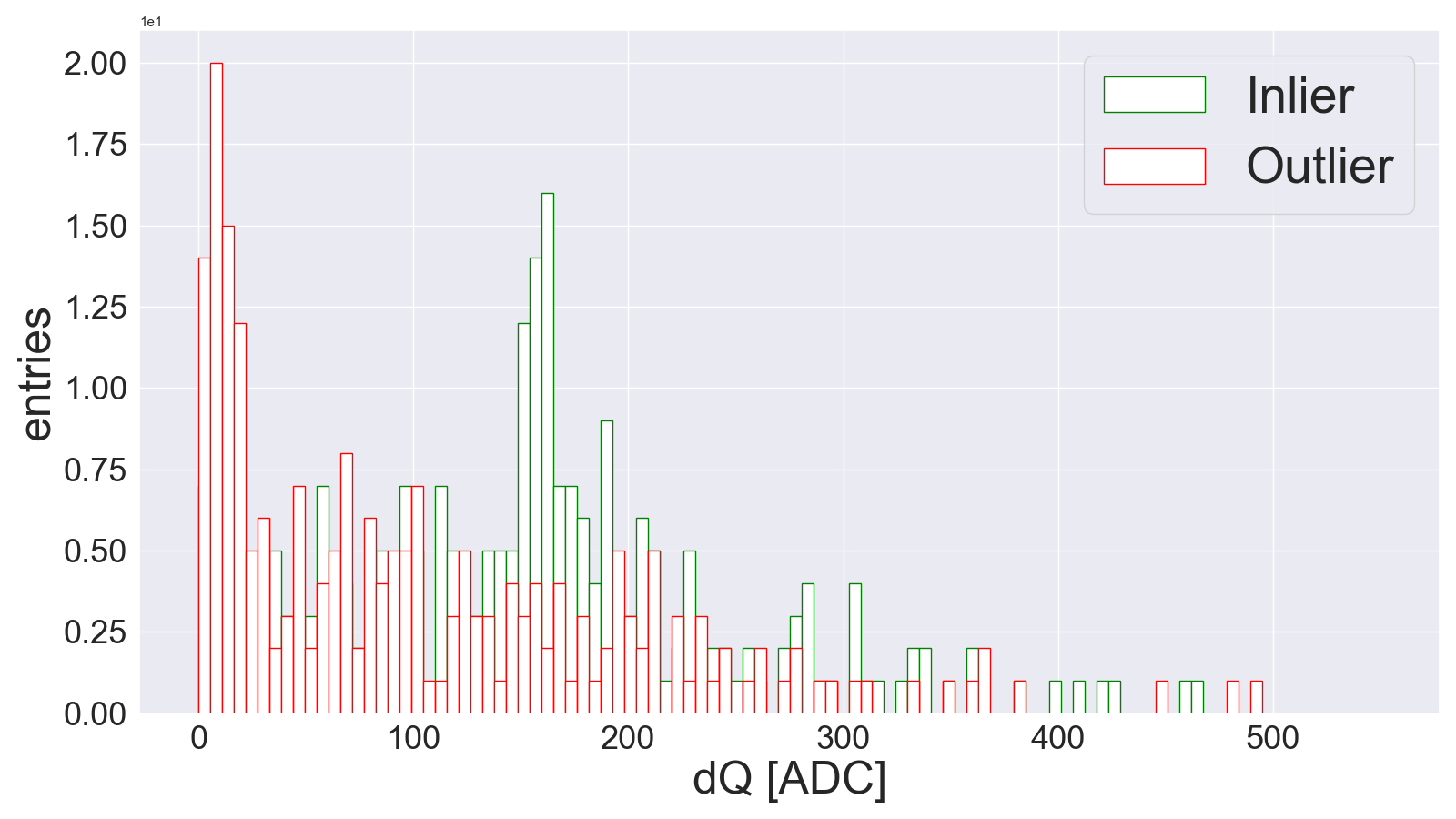}
        \caption{$dQ_{e}$ after processing} 
        \label{dQ_e}
    \end{subfigure}
    
\caption{Comparison of the muon and electron charge signals recorded in the rear telescope panel, shown before and after RANSAC processing. The charge is measured here after digitization of the energy deposits measured in GEANT4, and is given in ADC counts unit.}
\label{mu_e_signals}
\end{figure}

We compare in figure \ref{mu_e_signals} the muon and electron charge signals $dQ$ in the rear panel of our detector (bottom panel in figure \ref{evd_mu_cry}), located behind the lead shielding panel. We observe that 96\% of the muon signals on figure \ref{dQ_mu_ransac}, fitted with a Landau function, are tagged at as inlier. We also observe a suppression of noisy hits at lower ADC values. As for electrons, we observe a large proportion of outlier hits that populates the low ADC values, with a ratio outlier/inlier around 85\%.

\subsubsection{Angular resolution}
\begin{figure}[ht!]
    \begin{subfigure}[ht]{.45\textwidth}
      \centering
      \includegraphics[width=0.8\textwidth]{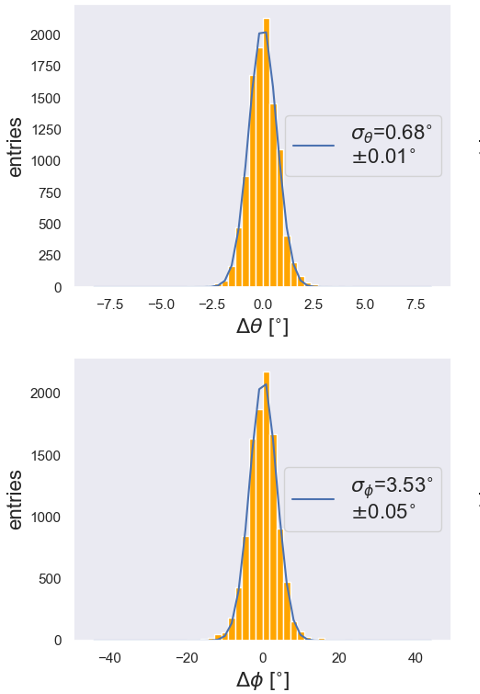}
      \caption{RANSAC}
      \label{res_ransac}
    \end{subfigure}%
    \hfill
    \begin{subfigure}[ht]{.45\textwidth}
      \centering
        \includegraphics[width=0.8\textwidth]{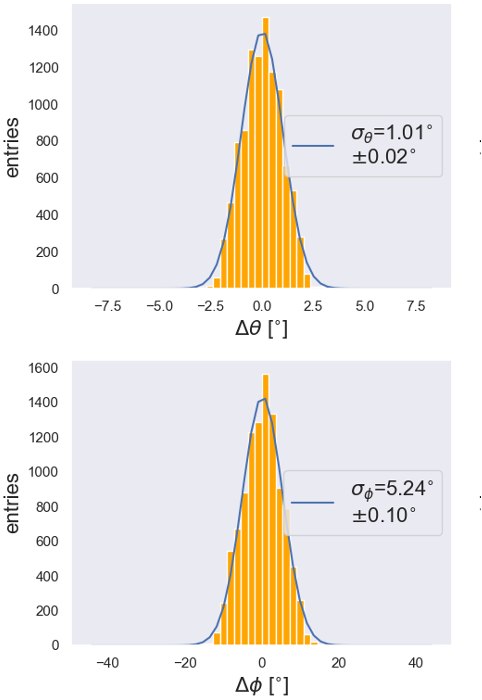}
        \caption{Straightness check} 
        \label{res_straight}
    \end{subfigure}
\caption{Angular resolution in Zenith (upper line) and Azimuth (bottom line) angles for the estimated muon trajectories, for the (\subref{res_ransac}) RANSAC and (\subref{res_straight}) and straightness check processing.}
\label{ang_res}
\end{figure}

In figure \ref{ang_res}, we show the angular deviation in zenith ($\Delta\theta$) and azimuth ($\Delta\phi$) between the generated muon straight track and the reconstructed track. We observe for the RANSAC processing an improvement of the angular resolutions in both zenith and azimuth of the order of 30\%, compared to the straightness check-based processing.

\subsubsection{Reconstruction efficiency}
\begin{figure}[ht!]
    \begin{subfigure}[ht]{.45\textwidth}
      \centering
      \includegraphics[width=0.8\textwidth]{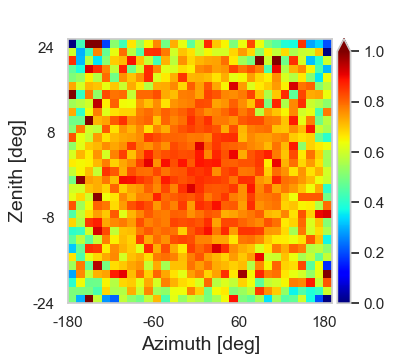}
      \caption{RANSAC}
      \label{eff_ransac}
    \end{subfigure}%
    \hfill
    \begin{subfigure}[ht]{.45\textwidth}
      \centering
        \includegraphics[width=0.8\textwidth]{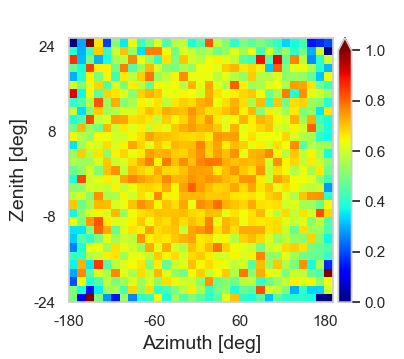}
        \caption{Straightness check} 
        \label{eff_straight}
    \end{subfigure}
\caption{Muon reconstruction efficiency $\epsilon_{\mu}$ maps for (\subref{eff_ransac}) RANSAC with a mean efficiency $\langle \epsilon_{\mu}\rangle=0.71\pm0.11$   and (\subref{eff_straight}) straightness check processing with $\langle \epsilon_{\mu}\rangle=0.61\pm0.10$. Each pixel is a given telescope line-of-sight.}
\label{eff}
\end{figure}

In figure \ref{eff}, we have measured the muon reconstruction efficiency $\epsilon_{\mu}$ for each 4-panel detector line-of-sight. For each pixel, representing a given line-of-sight, we show the ratio of the number of reconstructed tracks and the total number of generated muon straight track. We observe for the RANSAC processing an increase by 10\% of the mean efficiency $\langle \epsilon_{\mu}\rangle$ with respect to the straightness check-based processing.

\newpage

\section{ Application to an Open-Sky acquisition}
We present in this section the application of the RANSAC processing to real data taken with a 4-panels telescope, as described in \ref{secDetector}. The data covers 36 hours and has been recorded in open-sky mode, measuring the vertical cosmic muon flux centered around zenith $\theta=0^{\circ}$. This run serves as a calibration step for measuring the experimental detector acceptance, which allows to take into account detection matrices flaws due notably to some imperfect optical couplings \cite{marteau2016} in the dome-outgoing flux computation.
The integrated acceptance analytical function, presented in \cite{sullivan1971}, on the other hand, is purely geometrical and depends only on the dimensions of our detector: distance between front and rear matrices, number, length and width of scintillator bars in each matrices.  

\begin{figure}[ht!]
    \begin{subfigure}[ht]{.45\textwidth}
      \centering
      \includegraphics[width=\textwidth]{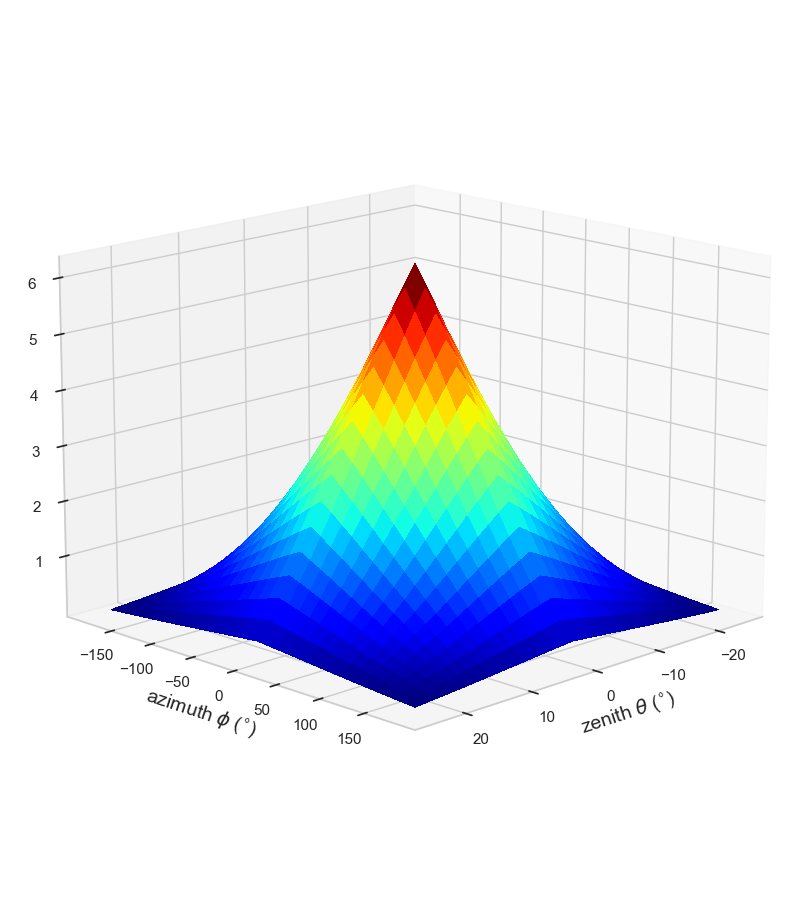}
      \caption{Theoretical acceptance}
      \label{theoretical_acc}
    \end{subfigure}%
    \hfill
    \begin{subfigure}[ht]{.45\textwidth}
      \centering
        \includegraphics[width=\textwidth]{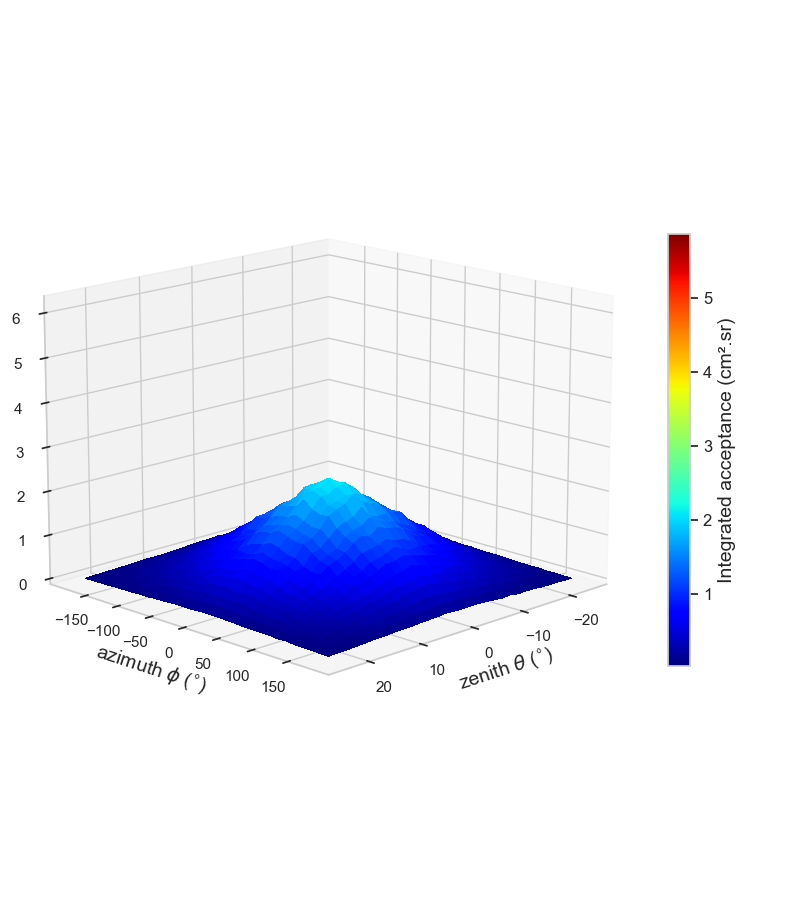}
        \caption{RANSAC experimental acceptance} 
        \label{experimental_ransac_acc}
    \end{subfigure}
    \caption{(\subref{theoretical_acc}) Theoretical integrated acceptance of the 4-panel telescope geometry and (\subref{experimental_ransac_acc}) experimental acceptance obtained after open-sky data RANSAC processing.}
\end{figure}

\begin{figure}[ht!]      
      \begin{subfigure}[ht]{0.5\textwidth}
        \includegraphics[width=\textwidth]{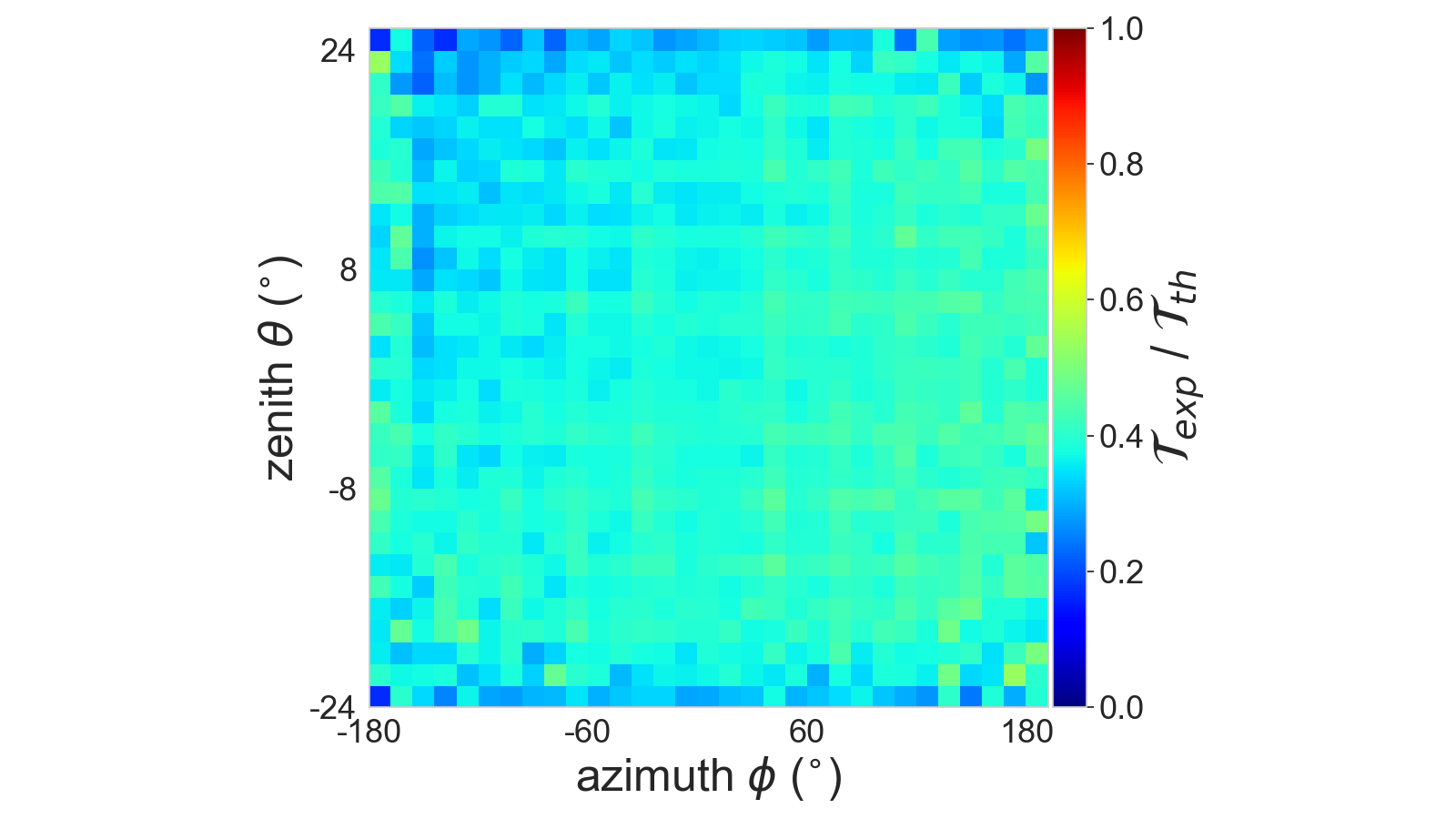}
        \caption{RANSAC}
        \label{ratio_acc_ransac}
       \end{subfigure}%
       \hfill
    \begin{subfigure}[ht]{0.5\textwidth}
      \includegraphics[width=\textwidth]{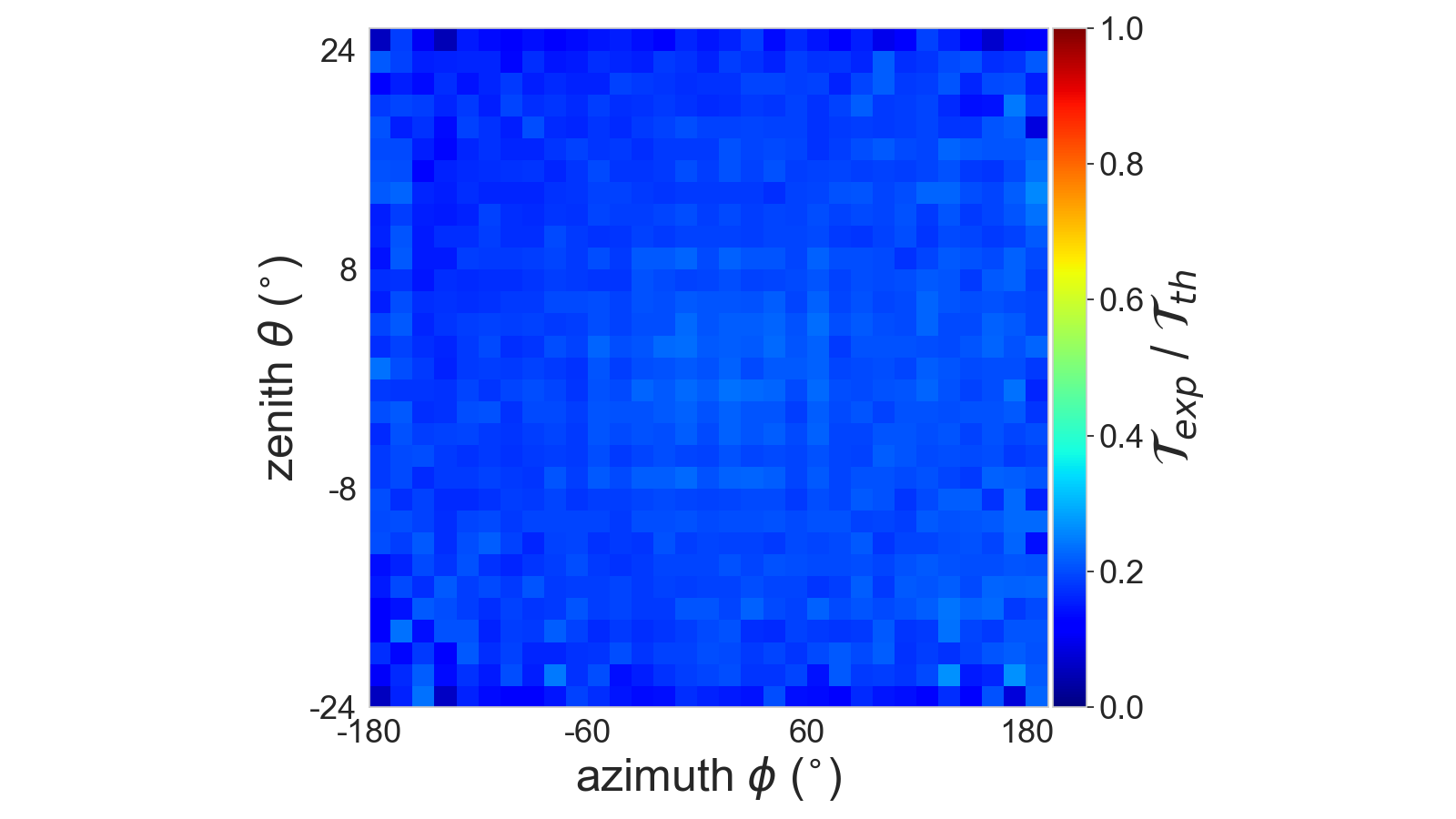}
        \caption{Straightness check}
        \label{ratio_acc_straight}
    \end{subfigure}%
      \caption{Ratios between experimental and theoretical integrated acceptance $\mathcal{T}_{exp}$/$\mathcal{T}_{th}$ for (\subref{ratio_acc_ransac}) RANSAC processing  and (\subref{ratio_acc_straight}) former straightness check processing .}
     \label{ratio_acc}
\end{figure}
We observe in figure \ref{ratio_acc} that the mean ratio between experimental and theoretical acceptances is $0.38\pm0.02$ using the RANSAC processing. This reduction factor is consistent with previous detection efficiency estimation obtained in \cite{lesparre2012}. As for the former processing algorithm, this ratio is equal to $0.19\pm0.01$, that is, reduced by a factor 2. This means that with the new RANSAC method we manage to reconstruct twice more events for real open-sky data. 
\section{{Summary and outlook}}
In this paper, we have presented a new reconstruction algorithm, based on the RANSAC procedure, for scintillator-based muon telescope data and compared it with our former processing algorithm that relied on a track straightness check criterion without any trajectory fit for the incident particles.
The main drawback of the RANSAC-based algorithm is in an increased processing time, multiplied by a factor 2 for $10^{5}$ MC muon events on a standard 8Go CPU.
In spite of this, we observe that a RANSAC-based reconstruction of the muon tracks in the detector offers significant advantages. From the RANSAC processing of a pure muon MC sample, we highlight :
\begin{itemize}
    \item a 30\% gain in muon angular resolution 
    \item a 10\% increase in muon reconstruction efficiency 
    \item a 13\% increase in the muon count
\end{itemize}
Although in a real world data acquisition, we would not be able to perform precise particle identification,  we have already observed for open-sky real data that the RANSAC-reconstructed event count is doubled compared to the one obtained with the former processing algorithm, leading to a telescope effective acceptance estimation closer to the theoretical integrated acceptance expected for the 4-panel telescope configuration.
This increase in reconstructed event rate will need to be further confirmed with a realistic CORSIKA\cite{corsika} cosmic flux generation, taking into account all the cosmic sources of physical noise, that were not present in the current GEANT4 simulation study.

\section{{Future studies}}
The RANSAC processing will be used to reconstruct the available tomography datasets taken both with new generation 4-panels telescopes presented in sec.\ref{secDetector}, and with former 3-panels telescopes.
The 2D density radiographies obtained for different scanned regions of the lava dome will then be combined to serve as input for numerical modelling, conjointly with other geophysics data sets (e.g, from gravimetry surveys), in order to obtain a 3D bulk density distribution model of the lava dome at an expected unprecedentedly high resolution thanks to the uniqueness of the different data sets available and the quality of the event reconstruction. This will lead to a better characterization of the hydrothermal fluid circulation impact on rock alteration and the resulting partial flank  collapse hazard.

\section*{Acknowledgements}
This work and the data shown were acquired in the context of the ANR MEGaMu.  We also wish to thank Axel Le Berrigaud, for his contribution to the first applications of the RANSAC-based reconstruction during his master thesis at IPGP, in 2020.

\bibliographystyle{unsrt}

\end{document}